# On Generation of Photons Carrying Orbital Angular Momentum in the Helical Undulator

A.Afanasev[*)], A.Mikhailichenko[**)]

[*)] Department of Physics, George Washington University, Washington, DC 20052, USA
[**)] CLASSE, Ithaca, NY 14853, USA

*Abstract.* We analyze properties of electromagnetic radiation in helical undulators with a particular emphasis on the orbital angular momentum of the radiated photons. We demonstrate that *all* harmonics higher than the first one radiated in a helical undulator carry an orbital angular momentum. We discuss some possible applications of this phenomenon and the ways of effective generation of these photons in a helical undulator. We call for review of results of experiments performed where the higher harmonics radiated in a helical undulator might be involved.

## OVERVIEW

Photons are powerful probes of the structure of matter. Depending on their wavelength, photons give us insight about phenomena at widely different scales, ranging from astrophysics to physics of elementary particles.

Motivated by the needs of nuclear and particle physics, here we will focus on angular momentum properties of electromagnetic radiation. The fact that circularly polarized plane-wave photons carry an angular momentum of $\hbar$ was demonstrated in a classical experiment [1]. Description of photons in terms of spherical waves and multipole expansion is also well established [2], [3] for radiative processes at atomic scales, with photons characterized by multiple units of angular momentum. Note, however, that separation of spin and orbital angular momentum of a photon cannot be done in a gauge-invariant way [4].

About 20 years ago, Allen and collaborators re-discovered[1] that a special type of beams (called Laguerre-Gaussian), predicted as non-plane wave solutions of Maxwell equations, can carry large angular momentum associated with their helical wave fronts [5]. When quantized, such beams can be described in terms of "twisted photons" [6]. Orbital angular momentum (OAM) light beams found numerous applications in optics, communications, biophysics, mechanics of micro-particles, and probes of Bose-Einstein condensates [7]. It was demonstrated [8] that in photo ionization of atoms by twisted photons the new selection rules apply that involve more than one unit of angular momentum.

Other possible application might be in the photo-cathodes for generation of highly polarized beams of electron with high brightness, where manipulation with the energy range of the photons carrying OAM might help in closing undesirable channels in emitting of photoelectrons from the levels prohibited by angular-momentum conservation laws. Recently, it was pointed out that twisted photons can be produced in MeV and GeV range via the mechanism of Compton backscattering [9].

---
[1] See ref [2], page 401, Appendix.

Beams of twisted photons with pre-set angular momentum may emerge as a new and productive tool in nuclear and particle physics with electromagnetic probes that would help to control angular momentum of particles or quantum states generated in photo-production processes. Applications may include: (a) production of high-spin states in the laser-based searches for dark matter particle candidates [10], [11]; (b) meson and baryon spectroscopy of high-spin states; (c) possible formation of new high-spin particles at high energies; (d) studies parity-violating effects in atomic transitions [12], [13]; e) searches for axion-like particles through analyzing polarization properties of light passing through strong magnetic fields [14].

This paper deals with a method of generation of OAM radiation based on a use of a helical undulator. Undulator radiation (UR) is a valuable source of radiation in the energy rage indicated. A lot of publications are dedicated to the UR properties [15]-[19]. One of the features of UR is that radiation is emitted in harmonics [15]

$$\omega_n = \frac{n\Omega}{1-\beta_\parallel Cos\vartheta},$$

where $n=1,2,3…$, enumerates the harmonics of frequency $\Omega = \beta_\parallel c/\bar{\lambda}_u$, $\lambda_u = 2\pi\bar{\lambda}_u$ is a spatial period of undulator field, $\bar{\beta} = \bar{v}/c \cong \beta_\parallel$, $\bar{v}$ is the particle's average longitudinal velocity in the undulator, and $\vartheta$ is the azimuthal angle with respect to the observer. Unfortunately, existing publications did not clarify the question about OAM carried by the photons radiated at harmonics higher that the first one, corresponding $n=1$. This is especially important for *helical undulators*, where a charged particle emits circularly-polarized photons. Although for the photon the orbital part of angular momentum and the spin part (=1) could not be separated, the peculiarities of generation of radiation in a helical undulator allow to distinguish the part of radiation which carries away the angular momentum of electron in an undulator; hence it is possible to distinguish with OAM. As we will demonstrate, all harmonics with $n >1$ carry OAM.

The physics explanation of the fact why the electron in a helical undulator generates radiation with OAM was not presented clearly in the literature.

In this paper we analyze undulator radiation in a helical undulator for its ability to generate the photons with OAM, and come to a simple and clear explanation why all harmonics other than the lowest one ($n=1$) carry OAM. This fact was *not* taken into consideration in the past in a number of experiments in atomic and nuclear physics obtained with the higher harmonics from helical undulators; see Ref. [20] for a few examples. Since spin polarization effects may be different if OAM photons are involved, re-analysis of relevant experiments may be in order.

## SOURCE OF ANGULAR MOMENTA

Angular momentum of radiation was described first in [22]. In [23], the concept of angular momentum of radiation was developed further. Most general description of the angular momentum was presented in [24]; see also [25]-[27], [30]. The angular momentum carried by EM radiation is defined as

$$\vec{M} = \frac{1}{c^2}\int \vec{r}\times(\vec{E}\times\vec{H})dV = \frac{1}{c^2}\int [\vec{E}\cdot(\vec{r}\cdot\vec{H}) - \vec{H}\cdot(\vec{r}\cdot\vec{E})]dV , \qquad (1)$$

where the integration is over the entire volume, the radius vector $\vec{r}$ is directed from the rotation axis to the point in a volume where the EM fields are located, see Fig.1 below. While the radius increases, the fields $\vec{E}, \vec{H}$ drop down, so the region near the instant position of the source gives the main contribution to the integral.

Since we are interested in a longitudinal component of angular momentum (directed along $z$), one can see, that the $M_z \neq 0$ only if $E_z \neq 0$ or $H_z \neq 0$,

$$M_z = \frac{1}{c^2}\int \vec{r}\times(\vec{E}\times\vec{H})\big|_z dV = \frac{1}{c^2}\int [E_z\cdot(\vec{r}\cdot\vec{H}) - H_z\cdot(\vec{r}\cdot\vec{E})]dV . \qquad (2)$$

The conditions for appearance of longitudinal component of light are described in Refs. [24], [27].

To explain appearance of longitudinal component in Ref.[24], the authors considered the EM wave with restricted transverse dimensions, demonstrating that the longitudinal component associated with derivative of the envelope function of EM field in a transverse direction.

For a helically-polarized EM wave there is no necessity for this, however, as the longitudinal component persists in this type of wave intrinsically while the point shifts from the axis. The easiest way to prove this is to consider the time dependent periodic EM field of the general nature. Without any restriction to the common case, this could be demonstrated with TM field[2]. This type of field could be defined by equations

$$\vec{\nabla}\times\vec{E}\big|_z = \frac{\partial E_y}{\partial x} - \frac{\partial E_x}{\partial y} = -\frac{\partial B_z}{\partial t} = 0 \qquad (3)$$

The components of electromagnetic field could be represented as following [31],

$$\bar{E} = E_x - iE_y = \frac{\partial W}{\partial u} + \frac{\partial W}{\partial \bar{u}}, \quad E_z = \int \left(\frac{\partial^2}{\partial z^2} - \frac{1}{c^2}\frac{\partial^2}{\partial t^2}\right)W(u,\bar{u},z,t)dz$$

$$B_z \equiv 0, \quad \bar{B} = B_x - B_y = i\varepsilon_0 \frac{\partial}{\partial t}\int \bar{E}(u,\bar{u},z,t)dz , \qquad (4)$$

where $u = x+iy$, $\bar{u} = x-iy$, $i^2 \equiv -1$, and $z$ is a longitudinal coordinate,

$$\frac{\partial}{\partial u} \equiv \frac{1}{2}\left(\frac{\partial}{\partial x} - i\frac{\partial}{\partial y}\right), \quad \frac{\partial}{\partial \bar{u}} \equiv \frac{1}{2}\left(\frac{\partial}{\partial x} + i\frac{\partial}{\partial y}\right). \qquad (5)$$

With the definition (5), Laplacian could be expressed as following

---

[2] The electric and magnetic photons corresponding to the representation of the field as TM and TE waves.

$$\nabla^2 \equiv \frac{\partial^2}{\partial x^2} + \frac{\partial^2}{\partial y^2} + \frac{\partial^2}{\partial z^2} = 4\frac{\partial^2}{\partial u \partial \bar{u}} + \frac{\partial^2}{\partial z^2}. \tag{6}$$

Complex potential $W(x,y,z,t) \equiv W(u,\bar{u},z,t)$ satisfies the equation

$$4\frac{\partial^2 W}{\partial u \partial \bar{u}} + \left(\frac{\partial^2}{\partial z^2} - \frac{1}{c^2}\frac{\partial^2}{\partial t^2}\right)W(u,\bar{u},z,t) = 0. \tag{7}$$

By introduction of potential $W(u,\bar{u},z,t) = \partial U/\partial z$, the longitudinal component of electric field from (4) can be represented as

$$E_z = \left(\frac{\partial^2}{\partial z^2} - \frac{1}{c^2}\frac{\partial^2}{\partial t^2}\right)U(u,\bar{u},z,t) = -4\frac{\partial^2 U}{\partial u \partial \bar{u}} \tag{8}$$

From the last expression one can see that any *plane* electromagnetic wave propagating in z-direction, $U(u,\bar{u},z,t) = U(z,t) = U(z-ct)$, does not have a longitudinal component, and, hence, does not accelerate particles moving along the straight line (z-coordinate) and does not carry angular momentum. One can see from Eq.(8) that the wave should have a transverse structure, $\partial^2 U/\partial u \partial \bar{u} \neq 0$ to be able to carry angular momentum (and accelerate particles) or to satisfy the equation $(\frac{\partial^2}{\partial z^2} - \frac{1}{c^2}\frac{\partial^2}{\partial t^2})U(u,\bar{u},z,t) \neq 0$. It could be satisfied if the potential can be factorized as $U(u,\bar{u},z,t) = \tilde{U}(u,\bar{u},z) \cdot f(z-ct)$, where $\tilde{U}(u,\bar{u},z) \equiv \tilde{U}(x,y,z) \equiv \tilde{U}(\vec{r})$ is a function of coordinates only.

Thus the source of the longitudinal component is associated with a nonzero derivative $\partial U/\partial \bar{u} \neq 0$. This property was used in Ref.[24] to demonstrate appearance of longitudinal component at the edge of a cylindrical EM beam. The other way to obtain a nonzero result from expression (8), as we claimed earlier, is helical symmetry of the EM wave. Indeed, the helical symmetry can be described as

$$W(\vec{r},t) = W(\vec{r},t) \cdot \exp(i2\pi z/\lambda) = W(u,\bar{u},z,t) \cdot [Cos(2\pi z/\lambda) + iSin(2\pi z/\lambda)], \tag{9}$$

where $\lambda$ stands for the spatial period along the coordinate z (wavelength of radiation). Formula (9) describes the helical field with *left helicity*, i.e. while the observer moves in a positive direction along z, the potential pattern rotates counterclockwise. Since the superposition rule is in effect here, the formula (9) can be treated as representation of two orthogonal polarizations. The electric field can be presented in the following form:

$$\bar{E} = E_x - iE_y = e^{i\frac{2\pi z}{\lambda}} \cdot \frac{\partial W}{\partial u} + e^{-i\frac{2\pi z}{\lambda}} \cdot \frac{\partial \overline{W}}{\partial u},$$

$$E_z = \operatorname{Re}\int\left(\frac{\partial^2}{\partial z^2} - \frac{1}{c^2}\frac{\partial^2}{\partial t^2}\right)W(u,\bar{u},z,t)e^{-\frac{2\pi z}{\lambda}} dz \tag{10}$$

Most important is the case of a dipole harmonic, which describes a helical undulator or a wiggler. In this case the only harmonics of interest are the ones having dipole symmetry. For a multipole harmonic the potential solution of (8) looks as follows [31]:

$$W(u,\bar{u},z,t) = (-i) \sum_{m=1}^{\infty} \frac{u^m}{m} \times$$

$$\times \left\{ G_{m-1}(z,t) - \frac{|u|^2}{4(m+1)} \left( \frac{\partial^2}{\partial z^2} - \frac{1}{c^2} \frac{\partial^2}{\partial t^2} \right) G_{m-1}(z,t) + \frac{|u|^4}{32(m+1)(m+2)} \left( \frac{\partial^2}{\partial z^2} - \frac{1}{c^2} \frac{\partial^2}{\partial t^2} \right)^2 G_{m-1}(z,t) - ... \right\} \quad (11)$$

where the functions $G_{m-1}(z,t)$ describe the value of multipole field at the axis $u=0$. If $G_{m-1}(z,t) = G_{m-1}(z - ct)$, then all terms in $\{\ \}$ brackets, except the first one, turn to zero, so that

$$W(u,\bar{u},z,t) = (-i) \sum_{m=1}^{\infty} \frac{u^m}{m} G_{m-1}(z-ct). \quad (12)$$

Substituting this expression into (10), one can obtain

$$E_z = \operatorname{Re} \sum_{m=1}^{\infty} \frac{u^m}{m} \int \left( \frac{\partial^2}{\partial z^2} - \frac{1}{c^2} \frac{\partial^2}{\partial t^2} \right) G(z-ct) e^{-i\frac{2\pi z}{\lambda}} dz = \operatorname{Re} \sum_{m=1}^{\infty} \frac{u^m}{m} \cdot G(z-ct) \cdot \frac{2\pi}{\lambda} \cdot e^{-i\frac{2\pi z}{\lambda}}. \quad (13)$$

In polar coordinates $u = re^{i\varphi}$ the latter can be expressed as

$$E_z = \operatorname{Re} \sum_{m=1}^{\infty} \frac{r^m}{m} \cdot G(z-ct) \cdot \frac{2\pi}{\lambda} \cdot e^{-i\frac{2\pi z}{\lambda} + im\varphi}. \quad (14)$$

One can see that the longitudinal component is equal to zero on the axis, but it is growing while the axis off-set is increasing. For example, the dipole helical longitudinal component of the field has a form $E_z = G(z-ct)\frac{r}{\lambda}\cos(\frac{z}{\lambda} + \varphi)$.

The simplest way to recognize the source of angular momenta is by considering a process of radiation in a reference frame moving with average velocity of the electron $\bar{v} = c\beta_\parallel = v\bar{\beta}$. In a moving frame the source of radiation is an electron orbiting along the circle with a radius $r'$, see Fig.1. We are interested in the radiation directed tangentially to the instant trajectory in this system, where $\vec{r} \perp (\vec{E} \times \vec{H})$, so the integral (1) can be evaluated as follows

$$\vec{M} = \frac{1}{c^2} \int \vec{r} \times (\vec{E} \times \vec{H}) dV \cong \int \frac{cdt}{c^2} \int \vec{r} \times (\vec{E} \times \vec{H}) dA = \frac{r'}{c} \int \left( \int \vec{S} dA \right) dt = \frac{r'}{c} \int I dt \quad (15)$$

where $dA$ stands for the element of area, $\vec{S} = \vec{E} \times \vec{H}$ is the Poynting vector, $I = \int \vec{E} \times \vec{H} dA = \int S dA$ is the intensity of radiation.
One can see that

$$\frac{d|\vec{M}|}{dt} = \frac{r'I}{c}. \quad (16)$$

On the other hand, for the electron losing its energy by synchrotron radiation (SR) and moving at a constant radius, the change of energy in a moving frame can be transformed as the following

$$I = \frac{dE}{dt} \cong c\frac{dp}{dt} = \frac{c}{r'}\frac{d(r'p)}{dt} = \frac{c}{r'}\frac{d|\vec{r}' \times \vec{p}|}{dt} = \frac{c}{r'}\frac{d|\vec{M}|}{dt} \; , \tag{17}$$

*i.e.* the loss of momentum of electron is equal to the momentum carried away by SR.

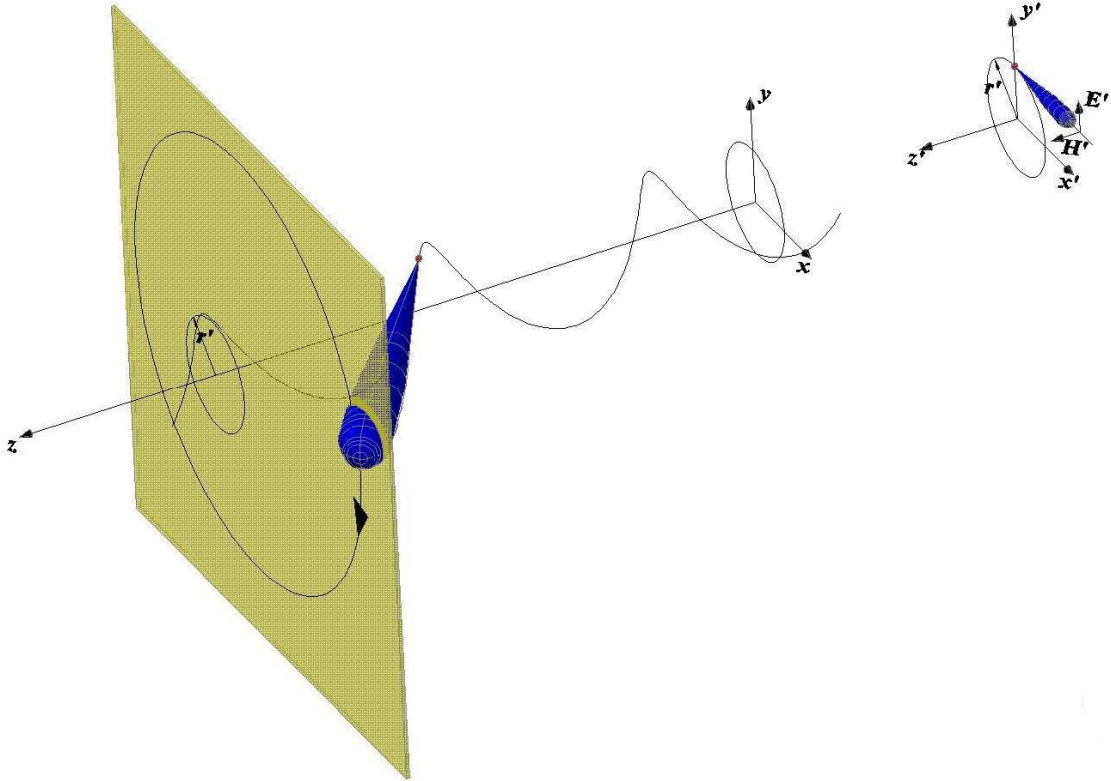

*Figure* 1. In a moving frame, the radiation has an electron as a point source moving along the circle with the radius $r' = \lambdabar_u K/\gamma$. In the Lab frame the cone of radiation is tilted toward the direction of motion (*z*-axis) by the angle $1/\gamma$, so the projector-type radiation is emitted from the off-axis location.

In formula (15) the integral $\int I dt$ is the total energy carried away by the radiation

$$\varepsilon^{tot} = \int I dt = \tfrac{2}{3} r_e^2 L_u H^2 \beta^2 \gamma^2 = \tfrac{2}{3} r_e^2 L_u \beta^2 \gamma^2 \left( \frac{2\pi m_e c^2 K}{e \lambda_u} \right)^2 = \frac{8\pi^2 e^2 K^2 L_u \beta^2 \gamma^2}{3\lambda_u^2} \tag{18}$$

where $K = eH\lambdabar/mc^2$ is a so-called undulator parameter or simply a *K*-factor. Total momentum carried away by the radiation comes to

$$p^{tot} = \frac{\varepsilon^{tot} v}{c^2} = \frac{8\pi^2 e^2 K^2 L_u \beta^3 \gamma^2}{3c\lambda_u^2}, \tag{19}$$

And the total angular momentum carried away by the radiation is

$$\Delta \vec{M} = m_e c [\vec{r'} \times \Delta \vec{p}] = m_e c [\vec{r'} \times \Delta \vec{p}_\perp] = r' K p^{tot} \vec{e}_z \tag{20}$$

We would like to emphasize here OAM carried away by higher harmonics as it is represented in Fig.1. (Note that the first harmonic is not shown in Fig.1, see Fig.2a). Since the intensity of radiation at the first harmonic is centered at the axis, it does not depend on the instantaneous position of electron, hence it does not carry OAM. The quantum number associated with the angular momentum can be represented as follows,

$$j = \frac{|\vec{M}|}{\hbar} = \frac{1}{\hbar c^2} \int r \times (\vec{E} \times \vec{H}) d\vec{r} \tag{21}$$

It can be seen in Fig.1 that since the source of radiation is located *at the off-axis position* (at the circle) the radiation is carrying away a linear momentum, and hence, *the angular momentum*. Simply speaking, the electron orbit is decreasing its radius.

One can see that by changing the *K* factor it is possible to change the angular momentum of radiation as more and more harmonics become involved in a process of radiation. The intensity of radiation carried away by the harmonics is described by the formula [15], [16], [19]

$$\frac{dI_n}{do} = \frac{e^2 n^2 \omega_0^2}{c(1-\beta_\| \cos\vartheta)^3} \left[ \beta_\perp^2 J_n'^2 \left( \frac{n\beta_\perp Sin\vartheta}{1-\beta_\| Cos\vartheta} \right) + \frac{(Cos\vartheta - \beta_\|)^2}{Sin^2\vartheta} J_n^2 \left( \frac{n\beta_\perp Sin\vartheta}{1-\beta_\| Cos\vartheta} \right) \right] \tag{22}$$

where $\bar{\beta} = \beta_\| \cong \beta \cdot (1 - \beta_\perp^2/2)$. One can see if $\beta_\| = 0$, then the formula (15) coincides with Schott's formula [33]. Graphs of $(c/e^2\omega_0^2)dI_n/do$ for three lower harmonics, $n=1$ and $n=2$ and $n=5$ for K=0.7 represented in Fig. 2.

In the coordinate system moving with the average velocity of electron, the electron moves by a circular trajectory with an instant radius $r' = \beta'\varepsilon'/eH$ (Fig.1). The intensity of radiation by the electron is given by summation of (22) over all harmonics and integration over the polar angle, so it gives again[3]

$$I = \frac{2}{3} \frac{e^4 H^4 v'^2 \gamma'^2}{m^2 c^5} = \frac{2}{3} c r_e^2 H^2 \beta'^2 \gamma'^2 \tag{23}$$

where the primes denote the quantities calculated in the moving system of reference: $v'$ is the electron's velocity, $\beta' = v'/c$, $\gamma' = \varepsilon'/mc^2 = 1/\sqrt{1-\beta'^2}$, $\varepsilon'$ is the electron's instant energy. Here, $r_e = e^2/mc^2$ is the classical electron radius. Note also that the radius $r'$ is an invariant under Lorentz transformation.

---

[3] Strictly speaking only the part of radiation with *n*>1 carries out the angular momentum, However, for *K*≥1 the intensity of radiation at the first harmonic is< 10%, so for evaluation this difference might be neglected.

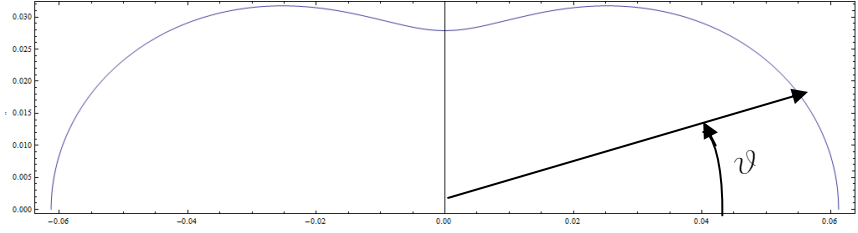

a) Harmonic **n**=1

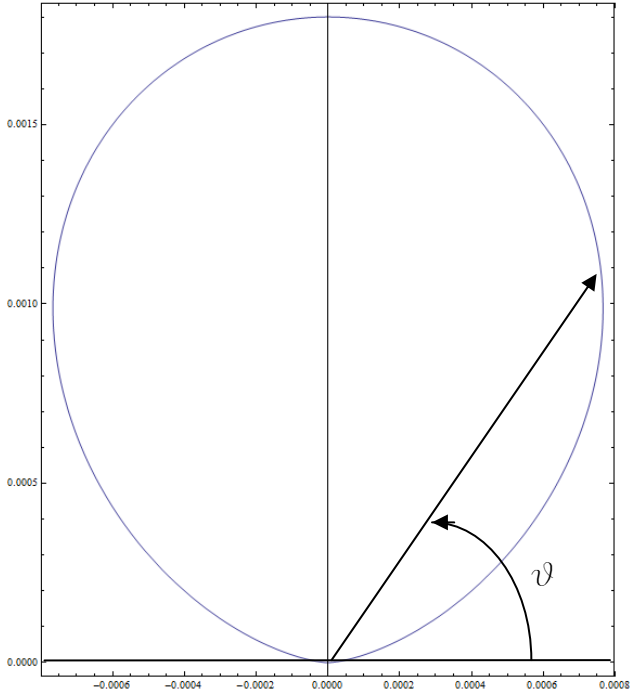

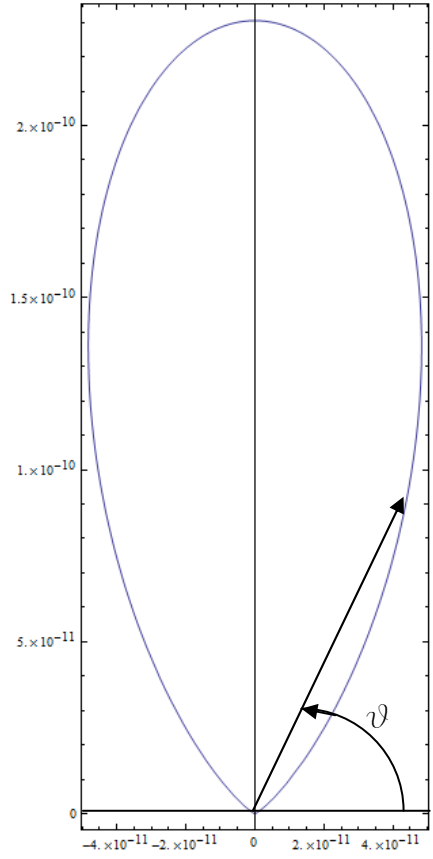

b) Harmonic *n*=2                          c) Harmonic *n*=5

*Figure* 2. The angular distribution of intensity of radiation on harmonics in a moving frame.

In a moving frame the electron is spinning along the circle with a radius $r' = \bar{\lambda}_u K / \gamma$, radiating SR within the cone which angular opening is defined by the local gamma-factor in the moving frame. So one can see that the source of radiation in a moving frame is located at this radius and so the Pointing vector associated with electron has orbital angular momentum with respect to the axis.

One peculiarity associated with helical motion in an undulator is that the orbital momentum of electron increases, while the energy of the electron drops down. Indeed, as the transverse momentum can be represented as

$$\left|\vec{M}\right| = r' \cdot p_\perp = r' \cdot mc\beta_\perp \gamma = r' \cdot mcK = mc\frac{\lambdabar_u K^2}{\gamma}, \quad (24)$$

where it was used the identity $K = \beta_\perp \gamma$ and the fact that $p_\perp$ and $r'$ are invariants under Lorentz transformations one can see that $\left|\vec{M}\right| \sim 1/\gamma$.

It is clear that the size of the source of undulator radiation can not be less, than the radius of helix $r' = \lambdabar_u K / \gamma$, what forces usage of beams with highest $\gamma$ and undulator with shortest period, operating at lower $K$. So for the ERL having $\gamma=10^4$, $K=1$ and period $\lambda_u = 2cm$, the minimal effective spot-size of the source will be not less, than $r' = \lambdabar_u K / \gamma \cong 1/3\ \mu m$, while for the LCLS this comes to ~1/6 $\mu m$. It is clear also, that operation with low K-factor should force operation with higher current to compensate the loss of the photon flux ~$K^2$.

## HELICAL UNDULATOR

Practically any helical undulator which is able to generate the field with $K \sim 1$ is suitable for generation of higher harmonics. But the preference is given to the ones with a possibility to change polarization, for example the SC one described in [28], allowing easy change of $K$ and polarization.

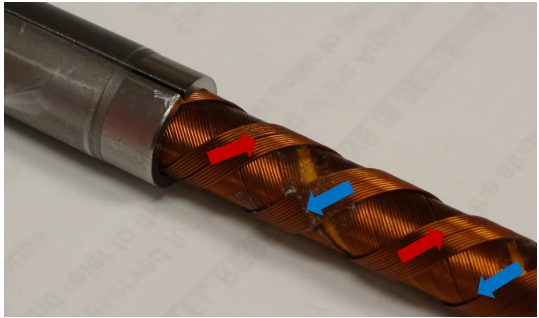

Figure 2. SC helical undulator with possibility to change the $K$ factor and polarization.

The coils in an undulator from Fig.2, wound above the Copper tube with six-wire strands stick together in flat SC cable. A period of winding is 24.5 *mm*. An outer diameter is 10 *mm*, and an inner diameter clear for the beam is Ø$_{inner}$=8 *mm*. The direction of current in the corresponding outer-layer coil is shown by the arrows [28]. With such aperture the SC undulator can generate $K$~1.5 for this period. For the larger period the achievable values of $K$ factor are growing exponentially for the fixed aperture.

# DISCUSSION

The fact that electromagnetic radiation from *any* helical undulator at higher harmonics carries the orbital angular momentum was not realized for many years.

The physics explanation of appearance of angular momenta for the photons radiated by electron in a helical undulator is clear, however. It is associated with the fact that the electron is moving along a helical trajectory twisted around the axis, and radiates while the point of radiation is shifted away from the axis.

Since in experiments with undulator radiation performed in the past, it was not realized that higher harmonics carry OAM, it is appropriate to review results obtained in such experiments, especially the ones for transitions between the states with different angular momenta.

Production of polarized positrons with higher undulator harmonics also needs revision. In the positron production scheme with undulator for ILC, the initial suggestion was to operate at low $K$-factor, $K<0.4$ [29], where the content of higher harmonics is minimal (even at $K\sim0.7$ only 50% of radiation is carried by higher harmonics). Meanwhile the baseline of ILC deals with undulator having $K\sim0.92$ [32] where content of higher harmonics is dominating. As the electron-positron pair should carry away the orbital momentum, their energies should be close to each other. It means that the energy of positron (electron) should be about ~half of the energy of quanta within narrow margins. Therefore the polarized positron production by higher undulator harmonics should be suppressed by the phase volume and by polarization behavior, as the positrons with half energy carry ~50% polarization at most.

# SUMMARY

The radiation in a helical undulator at higher harmonics is dominantly carrying orbital angular momentum, which should be taken into account. Physical meaning of this phenomenon is clear: the radiating electron moves along the helix of nonzero radius, so the source of instantaneous radiation is shifted from the axis. This fact was ignored in all experiments done with helical undulators in the past. We are suggesting revision of all results obtained with usage of higher undulator harmonics generated in a helical undulator. In addition, the theory of OAM photon interaction in matter (ionization, Compton scattering and electron-positron pair production) also requires revision.

The definition of brightness (brilliance) for the sources with helical undulators should be modified also.

In conclusion we note that usage of photons with OAM might open new ways in obtaining highly polarized electrons from photocathodes.

One of Authors (AM) thanks Evgenyi Bessonov for useful discussions.